# Comment on "On the negative value of dielectric permittivity of the water surface layer" [Appl. Phys. Lett. 83, 4506 (2003)]

G. Annino[a] and M. Cassettari
*Istituto per i Processi Chimico-Fisici del CNR, via G. Moruzzi, 1, 56124 Pisa, Italy*
(January 8, 2004)

In a recent letter [1] Cherpak et al. reported some measurements on the frequency shift and the merit factor of a whispering gallery dielectric resonator loaded by water and other materials. The obtained results appear of interest, since the dielectric properties of the samples can be potentially determined. In the analysis of the experimental data there is however a basic error. The authors claim that the sign of the shift of the resonance frequency evidences a negative dielectric constant of the water surface layer; this conclusion is wrong, since the effects of the imaginary part of the dielectric constant of the water are neglected.

The complex dielectric constant of water at 25 C and 34.88 GHz is given by $\hat{\varepsilon}_w = \varepsilon' - i \cdot \varepsilon'' = 23 - i \cdot 31.5$ [2]. The imaginary part is higher than the real one, and much higher than the dielectric constant of the material forming the resonator; it is then expected to give a sensible influence on the electromagnetic field distribution.

The effect of a high imaginary component of $\hat{\varepsilon}_w$ can be understood from the macroscopic Maxwell equations, in which the contribution of free charges and the absorption bands of bound charges are both encoded in $\varepsilon''$, so that similar effects are expected. In particular, when the $\varepsilon''$ of a dielectric becomes higher and higher in comparison to $\varepsilon'$, the effect on the electromagnetic field distribution becomes closer and closer to that one of a conductor. In this respect it is correct to compare the influence of the water to that of a metal for which, in the investigated frequency range, $|\varepsilon''| \gg |\varepsilon'|$ according to the Hagen-Rubens limit [3].

For the resonance mode under analysis, which has dominant axial electric field $E_z$ and radial magnetic field $H_\rho$ (Transverse-Magnetic mode, TM), a metal-like sample is expected to push out the fields, due to the boundary conditions, and then to raise the resonance frequency. Analogously, for Transverse-Electric (TE) modes a metal-like sample is expected to pull in the fields, and then to lower the resonance frequency.

In order to verify this interpretation, the different configurations of Ref. 1 have been theoretically analyzed by means of the modelling described in Ref. 4, which allows accurate calculations of complex resonance frequency and field distribution, also in presence of lossy regions. For the experimental conditions of Ref. 1, the resonance frequency shift and the merit factor variation of the mode $TM_{36,1,0}$ have been calculated; the obtained results are shown in Fig. 1.

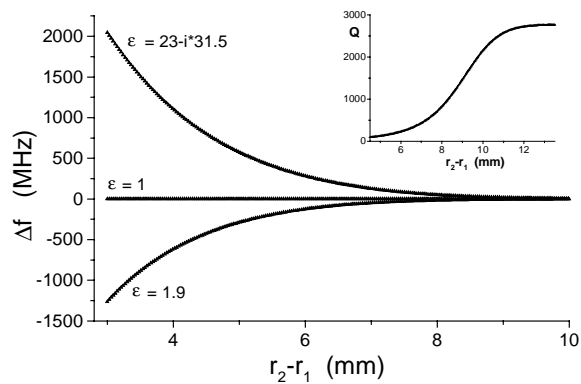

Fig.1 Calculated frequency shift and Q merit factor vs $r_2-r_1$ in the same conditions of Ref. 1.

All the basic features reported in Ref. 1 are well reproduced by using the correct complex dielectric constants. Specific field distributions are shown in Fig. 2.

---

[a] Electronic mail: geannino@ipcf.cnr.it

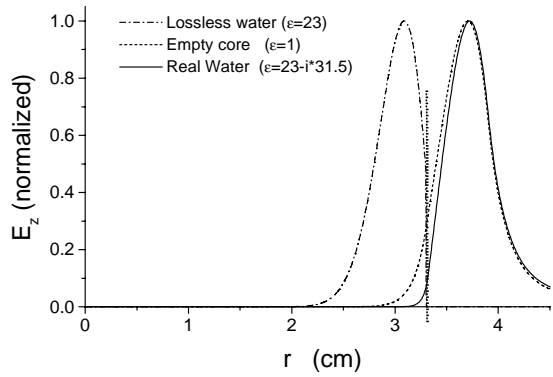

Fig.2 Calculated axial electric fields along the radius r of the resonator. Outer radius of the resonator $r_2$=3.9 cm, inner radius $r_1$=3.3 cm. The radius $r_1$ is marked with the dashed-dotted line. For details see Ref. 1.

The field deformation due to the dielectric losses of water is evident.

The same analysis has been done on some quasi-TM and quasi-TE modes with $s \neq 0$; as expected, the frequency shift due to water results positive in quasi-TM modes and negative in quasi-TE modes. These predictions have been also experimentally verified on quasi-TM and quasi-TE modes of a whispering gallery dielectric resonator without metallic shielding.

In conclusion, the experimental results reported in Ref. 1 are well reproduced without invoking a negative dielectric constant for water. A positive shift of the resonance frequency of quasi-TM modes is expected in this geometry for any sample having extremely high dielectric losses.

[1] N.T. Cherpak, A.A. Barannik, Yu.V. Prokopenko, T.A. Smirnova, and Yu.F. Filipov, Appl. Phys. Lett. **83** (22), 4506, (2003)
[2] W.J. Ellison, K. Lamkaouchi, and J.-M. Moreau, J. Mol. Liq. **68**, 171 (1996)
[3] F. Wooten, *Optical Properties of Solid* (Academic Press, New York, 1972)
[4] G. Annino, M.Cassettari, M. Martinelli, and P.J.M. van Bentum, Appl. Magn. Reson. **24**, 157 (2003)